# Production of W bosons in p-Pb collisions measured with ALICE at the LHC

**Edith Zinhle Buthelezi[1]**

*for the ALICE Collaboration*

*iThemba Laboratory for Accelerator Based Sciences*
*Old Faure Road, Faure, 7131, Western Cape, South Africa*
*E-mail: edith.zinhle.buthelezi@cern.ch*

W bosons, which are produced in hard scattering processes of partons in collisions of hadrons, do not interact strongly with the medium produced in high-energy heavy-ion collisions. Therefore, in p-Pb collisions the measurement of W-boson yields represents a standard candle to check the validity of binary scaling and can provide important constraints on the parton distribution functions, which can be modified in nuclei with respect to protons or neutrons.

At the LHC, ALICE (A Large Ion Collider Experiment) is dedicated to the study of ultra-relativistic heavy-ion collisions, in which a hot and dense, strongly-interacting medium is formed. At forward rapidity ALICE is equipped with a muon spectrometer that allows measurements of dimuon decays of quarkonia, muons from heavy-flavour hadron decays and also W bosons via their single-muon decay. In ALICE W-boson cross sections were measured in p-Pb collisions at $\sqrt{s_{NN}} = 5.02$ TeV via the contribution of their muonic decays to the inclusive $p_T$-differential muon yield measured at forward ($2.03 < y^{\mu}_{cms} < 3.53$) and backward ($-4.46 < y^{\mu}_{cms} < -2.96$) rapidity. Recent results obtained from these measurements are presented and compared to perturbative Quantum Chromodynamics calculations at next-to-leading order.

---

[1]Speaker

1. Introduction

Electroweak W bosons were discovered at the CERN Super Proton Synchrotron (SPS) [1]. They are massive particles (80.385 ± 0.015 GeV/$c^2$) produced in initial hard parton scattering processes with a formation time of approximately 0.003 fm/$c$ and have a lifetime of about 0.09 fm/$c$ [2]. Their properties have been studied extensively at hadron colliders [3]. In proton-proton collisions their cross sections are known with a precision limited by the parton distribution function (PDF) uncertainties. The measurements of W bosons via the leptonic decay channels (branching ratios of 10.57 ± 0.15%) are not affected by the presence of strongly-interacting matter. The unprecedented energies available at the Large Hadron Collider (LHC) make it possible to measure W bosons in pp, Pb-Pb and p-Pb collisions and will allow to probe parton distribution functions and their nuclear modification at Bjorken-$x$ ranges $x \sim (10^{-4} - 10^{-1})$ at large scales $Q^2 = M_W/2$ [4]. In p-Pb collisions the measurements of their yields allow the investigation of cold nuclear matter effects and constrain nuclear parton distribution functions (nPDF) [5][6]. Also, they serve as an important baseline for the understanding and the interpretation of Pb-Pb data.

In this report, preliminary results are presented obtained with ALICE from the study of W-boson production in the single muon decay channel, $W^\pm \to \mu^\pm \nu$, at forward ($2.03 < y^\mu_{cms} < 3.53$) and backward ($-4.46 < y^\mu_{cms} < -2.96$) rapidity for muons with transverse momentum $p^\mu_T > 10$ GeV/$c$ in p-Pb collisions at $\sqrt{s_{NN}} = 5.02$ TeV. The measurements are compared to predictions from pQCD calculations at next-to-leading-order (NLO) [5] and are complementary to results recently published by the CMS Collaboration [7].

2. Experimental setup and data sample

As the detailed description of the ALICE detector is given in [8][9]here parts of the detector used in the present measurements are briefly described. ALICE comprises of central barrel detectors embedded in a solenoid magnet (B = 0.5 T) covering the pseudo-rapidity of |$\eta$| < 0.9 where hadrons, electrons and photons are measured, and a forward muon spectrometer. Information on global event characterization is provided by the VZERO detectors, which are two plastic scintillators, V0A and V0C, situated on either side of the interaction point at $2.8 < \eta < 5.1$ and $-3.7 < \eta < -1.7$, the two layers of the Silicon Pixel Detector (SPD) located in the central barrel and the Zero Degree Calorimeters, ZNA and ZNC, located at a distance of 112 m on either side of the interaction point along the beam pipe. The VZERO detector is also used for triggering as well as for offline rejection of beam-gas events while the SPD is also used to determine the interaction vertex. The muon spectrometer, located at $-4.0 < \eta < -2.5$, is composed of a passive front absorber of 10 interaction lengths (10 $\lambda_{int}$) designed to filter hadrons, photons, electrons and muons from light hadron decays before the spectrometer, a dipole magnet providing a field integral of 3 Tm, five tracking stations each composed of two planes of Cathode Pad Chambers, a muon filter which absorbs most of the punch-through hadrons and lastly, the two trigger stations equipped with two planes of Resistive Plate Chambers located downstream of the tracking system.

The analysis is based on data collected in 2013 in p-Pb collisions at $\sqrt{s_{NN}} = 5.02$ TeV. Due to the LHC design, the colliding beams have different energies per nucleon: $E_p$ = 4 TeV, $E_{Pb}$ =

1.58 $A_{Pb}$TeV ($A_{Pb}$ = 208 is the mass number of the Pb nucleus). Consequently, the centre-of-mass of the nucleon-nucleon collision is shifted by $\Delta y = 0.465$ with respect to the laboratory frame in the direction of the proton beam. Data were taken in two configurations by inverting the orbits of the two beams. In this way, both forward ($2.03 < y^{\mu}_{cms} < 3.53$) and backward ($-4.46 < y^{\mu}_{cms} < -2.96$) nucleon-nucleon centre-of-mass rapidities are covered, with the positive rapidity defined by the direction of the proton beam. The integrated luminosities for p-Pb and Pb-p collisions are 4.9 nb$^{-1}$ and 5.8 nb$^{-1}$, respectively. The data sample consists of minimum bias trigger events (MB - coincidence of a signal in both the event counters and the VZERO detectors) and tracks in the muon trigger system with $p_T \geq 4$ GeV/$c$. An offline selection is applied and events without a reconstructed primary vertex from the SPD are rejected. Furthermore, tracks are required to be reconstructed within the acceptance of the muon spectrometer: $-4.0 < \eta_{lab} < -2.5$, $170° < \theta_{abs} < 178°$, where $\theta_{abs}$ is the polar angle at the end of the absorber. To eliminate tracks from punch-through hadrons each track candidate of interest in the tracking system is required to match the corresponding track reconstructed in the trigger system. Finally, a $p$DCA cut (correlation of the momentum, $p$, and the distance of closest approach, DCA) is applied within 6$\sigma$, where $\sigma$ is the standard deviation of the distribution in order to remove fake, beam-gas tracks and particles produced in the absorber.

3. Data analysis

In the analysis the W-boson signal, i.e. the number of muons from the decays of W bosons, $N_{\mu \leftarrow W}$, is extracted by taking into account the contributions of W$^+ \to \mu^+ \nu_\mu$ and W$^- \to \mu^- \bar{\nu}_\mu$ to the inclusive single-muon $p_T$ distributions. The decay kinematic provides two final state particles with $p_T \sim M_W/2 \sim 40$ GeV/$c$. Since the neutrino cannot be detected, the signature is a high-$p_T$ muon with a large missing transverse energy and it is characterized by a $p_T$ distribution which dominates inclusive muon yield at $p_T > 30$ GeV/$c$, with a Jacobean peak at $p_T \sim 40$ GeV/$c$. The main sources of background contributing to the inclusive $p_T$ spectra are semi-muonic decays of heavy-flavour (charm and beauty hadrons) which dominate at $10 < p_T < 35$ GeV/$c$ [4] as well as dimuons from Z$^0/\gamma^*$ decays which populate the region of $p_T > 50$ GeV/$c$. The yield of muons from W-boson decays is extracted by fits based on suitable parameterizations of the different components. The components of W and Z-boson decays are described by using [10] based on next-to-leading-order Monte Carlo (MC) event generator templates [10] with CT10 [14] PDF and PYTHIA [11] to take into account the [12] nuclear modification of the PDF. The isospin is accounted for in the template by generating pp and pn collisions and combining the results using

$$\frac{dN_{pPb}}{dp_T} = \frac{Z}{A} \frac{1}{N_{pp}} \frac{dN_{pp}}{dp_T} + \frac{A-Z}{A} \frac{1}{N_{pn}} \frac{dN_{pn}}{dp_T} \qquad (1)$$

A = 208 and Z = 82 are mass and atomic number of the Pb nucleus, respectively. The different contributions are taken into account and summed together in the final fit function defined by

$$f(p_T) = N_{bkg} f_{bkg}(p_T) + N_{\mu \leftarrow W} f_{\mu \leftarrow W}(p_T) + N_{\mu \leftarrow Z/\gamma^*} f_{\mu \leftarrow Z/\gamma^*}(p_T) \qquad (2)$$

where $f_{bkg}(p_T)$ represent templates for muons from heavy-flavour decays based on a pQCD calculation at fixed-order-next-to-leading-log (FONNL[13], $f_{\mu \leftarrow W}(p_T)$ and $f_{\mu \leftarrow Z/\gamma^*}(p_T)$ are templates for W and Z/$\gamma^*$ components based on POWHEG [10]. The free parameters are $N_{bkg}$ and $N_{\mu \leftarrow W}$ which are the number of muons from heavy-flavour decays and W bosons, respectively, while $N_{\mu \leftarrow Z/\gamma^*}$, i.e. the number of muons from Z/γ* decays, is not a free parameter in the fit; it is fixed to the number of muons from W decays, $N_{\mu \leftarrow W}$. Furthermore, the signal is corrected for acceptance and efficiency using POWHEG [10] simulations. Finally, $N_{\mu \leftarrow W}$ is normalized to the number of minimum bias (MB) events. The systematic uncertainties associated with the signal extraction account for ~6 - 24%, tracking and trigger efficiency ~2.5%, detector alignment accounts for ~1%, normalization to MB events ~7.2% and up to 8.6% to account for pile-up effects.

4. Results

Preliminary results for the production cross sections, $\sigma_{\mu \leftarrow W}$, of muons from W-boson decays obtained in this analysis are presented in two rapidity intervals, forward ($2.03 < y^{\mu}_{cms} < 3.53$) and backward ($-4.46 < y^{\mu}_{cms} < -2.96$) in Figure 1. The isospin effects are clearly visible, i.e. at backward rapidity (Pb ion beam direction) an enhancement is seen for W⁻ where $\sigma_{\mu^- \leftarrow W^-}$ is larger than $\sigma_{\mu^+ \leftarrow W^+}$ whereas at forward rapidity $\sigma_{\mu^- \leftarrow W^-}$ is approximately equal to $\sigma_{\mu^+ \leftarrow W^+}$. This could potentially arise from chirality, i.e. the production of W⁺ is suppressed because the proton beam is coming from the spectrometer side (negative rapidity) and W⁺ is boosted towards the direction of the proton. In Figure 2 the same results are compared with POWHEG [10] predictions which do not include nuclear shadowing effects (nPDF). Predictions by POWHEG [10] slightly over-estimate $\sigma_{\mu^+ \leftarrow W^+}$. For $\sigma_{\mu^- \leftarrow W^-}$ the agreement is good. In Figure 3 and Figure 4 we compare the measured cross sections with pQCD theoretical predictions [5] at next-to-leading order (NLO) assuming both unmodified (CT10 [14]) and modified (CT10+EPS09 [12] nPDFs. Taking into account the EPS09 prescription of nuclear shadowing of the PDFs further improves the agreement between data and theory, especially at the forward rapidity. These results are also consistent with observations made by the CMS collaboration [7].

5. Summary

We have measured W-boson production via the single-muon decay channel at forward and backward rapidity in p-Pb collisions at $\sqrt{s_{NN}}$ = 5.02 TeV. The comparison of experimental results with theoretical calculations shows a good agreement within uncertainties. Further improvements are seen when taking into account the EPS09 prescription of nuclear shadowing of the PDFs. These results are consistent with what has been observed by the CMS Collaboration [7] and they complement the rapidity range covered at the LHC. The measurements will be useful to further constrain pQCD calculations as suggested in [5].

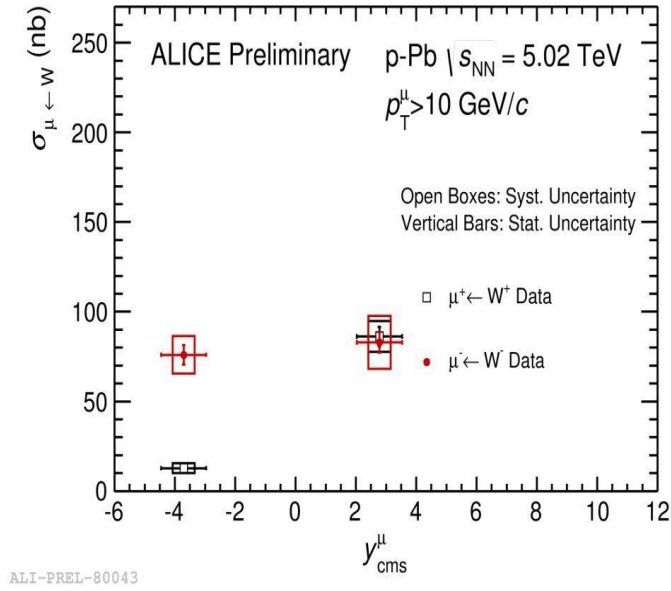

Figure 1 : Cross sections of W- boson production at forward and backward rapidity in p-Pb collisions at $\sqrt{s_{NN}}$ = 5.02 TeV.

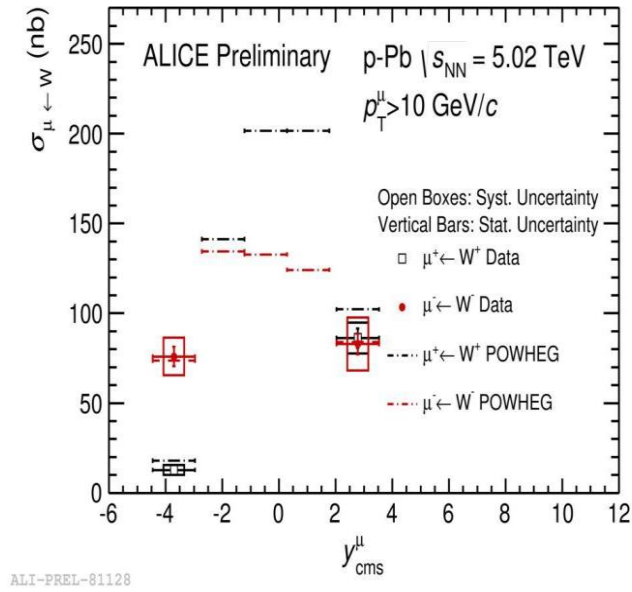

Figure 2 : Cross sections of W-boson production at forward and backward rapidity in p-Pb collisions at $\sqrt{s_{NN}}$ = 5.02 TeV compared with predictions by [10]

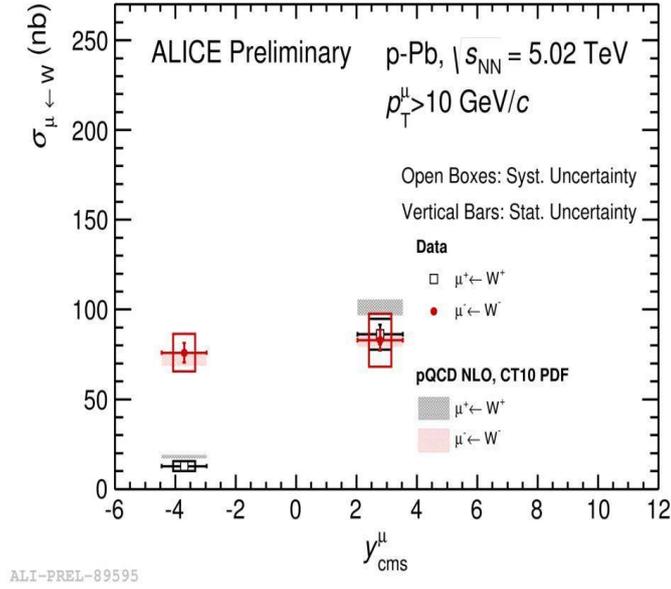

Figure 3 : Cross sections of W-boson production at forward and backward rapidity in p-Pb collisions at $\sqrt{s_{NN}}$ = 5.02 TeV compared with theoretical predictions based on pQCD NLO with CT10 [14].

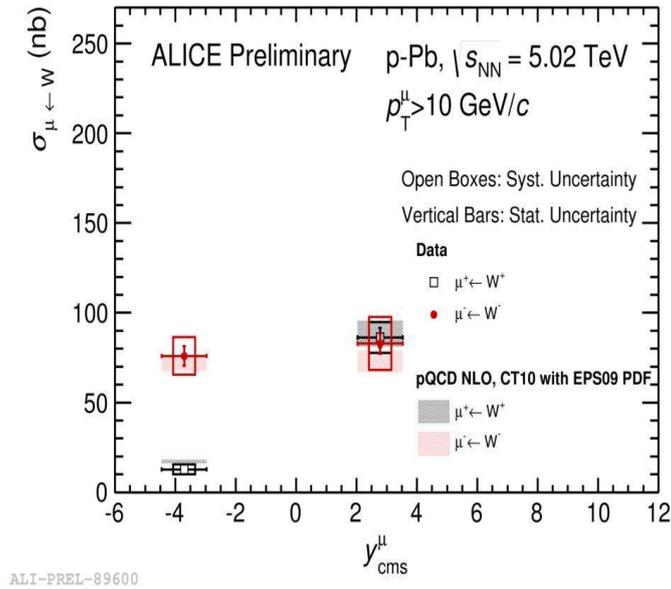

Figure 4 : Cross sections of W-boson production at forward and backward rapidity in p-Pb collisions at $\sqrt{s_{NN}}$ = 5.02 TeV compared with theoretical predictions based on pQCD NLO with CT10 [14] and EPS09 PDF from [12].

**Acknowledgements**

The author would like to thank the Department of Science and Technology of South Africa, the National Research Foundation and iThemba LABS for financial support.

## References

[1] G. Arnison et al., *Phys. Lett. B 122 (1983) 103*.

[2] J. Beringer, *et al*., *Phys. Rev.* D 86 (2012) 010001.

[3] W-M Yao*, et al., J. Phys.* G 33 (2006) 1.

[4] Z Conesa del Valle, *Eur. Phys. J. C6 (2009) 729-733*.

[5] Hannu Paukkunen and Carlos A. Salgado, *JHEP* 1103 (2011) 071.

[6] Peng Ru, Ben-Wei Zhang, Luan Cheng, Enke Wang, Wei-Ning Zhang, J. *Phys. G: Nucl. Part. Phys. 42 (2015) 085104*.

[7] The CMS Collaboration, *arXiv: 1503.05825 [nucl-ex]*.

[8] The ALICE Collaboration, *J. Instrum. 3, (2008) S08002*.

[9] The ALICE Collaboration, *Int. J. Phys. Z 29 (2014) 1430044*.

[10] S. Alioli, P. Nason, C. Oleari, E. Re, *JHEP 0807 (2008) 060*.

[11] Torbjörn Sjöstrand, Stephen Mrenna and Peter Skands, *JHEP 05 (2006) 026*.

[12] K.J. Eskola, H. Paukkunen and C.A. Salgado, *JHEP 0904 (2009) 065*.

[13] M. Cacciari, S. Frixione, N. Houdeau, M.L Mangano, P Nason, et al., *JHEP 1210 (2012) 137*.

[14] Hung-Lai Lai, Guzzi Marco *et al*., *Phys. Rev. D82 (2012) 074024*.